\documentclass[
aps,nofootinbib,
showpacs,showkeys,preprint
tightenlines,preprintnumbers,] {revtex4}

\usepackage{epsf,epsfig,subfigure,graphicx,amsmath,amssymb 
}
\usepackage{color} 
\newcommand{\ie}{{\it i.e.~}}
\newcommand{\etal}{{\it et al.\,}}

\newcommand{\Qem}{Q_{\rm em}}

\newcommand{\gev}{\,\textrm{GeV}}

\newcommand{\Mg}{{M_{\rm GUT}}}

\def\sw0{{$\sin^2\theta_W^0$}}

\def\E6{{\rm E_6}}

\def\EE8{{\rm E_8\times E_8'}}

\def\oneep{{\varepsilon}}
\def\halfep{{\sqrt{\varepsilon}}}
\def\thalfep{{{\varepsilon}^{3/2}}}

\begin{document}

\draft

\title{\Large\bf  On the progenitor quark mass matrix
}

\author{Jihn E.  Kim $^{(1,2)}$ and Se-Jin Kim$^{(1)}$}
\address{$^{(1)}$Department of Physics, Kyung Hee University, 26 Gyungheedaero, Dongdaemun-Gu, Seoul 02447, Republic of Korea, 
 $^{(2)}$Department of Physics and Astronomy, Seoul National University, 1 Gwanakro, Gwanak-Gu, Seoul 08826, Republic of Korea
}

\begin{abstract} 
\noindent
We determined the quark mass matrix in terms of a small expansion parameter $\halfep$, which gives correctly all the quark masses and the CKM matrix elements at the electroweak (EW) scale, and obtain a progenitor form at the GUT scale by running the EW scale mass matrix.  Finally, a possible texture form for the progenitor quark mass matrix is suggested.

\end{abstract}
\maketitle

\section{Introduction}\label{sec:Introduction}

\bigskip
\noindent
The huge  hierarchy on the quark masses is one of the most important flavor problems in the Standard Model (SM) of particle physics. The charged current (CC) weak interactions in the  SM are well established, which are summarized by the  Cabibbo--Kobayashi--Maskawa (CKM) and Pontecorvo--Maki--Nakagawa--Sakata (PMNS) matrices  \cite{Cabibbo63,KM73,PMNS1,PMNS2,PMNS3}. The CKM  matrix  being close to the identity is due to this huge hierarchy \cite{FK19}. But neutrino masses do not reveal a hierarchy and hence the PMNS matrix need not be close to the indentity. 

\bigskip
\noindent
Since the strange quark has been found earlier than the charm quark, the mass hierarchy between  down-type (\ie  $\Qem= -\frac13$ quarks)  quarks has been usually used \cite{GOR68}. This continued in the gauge theory era.  Weinberg's finding of the accidental coincidence of the Cabibbo angle $\sin\theta_C$ being close to $\sqrt{m_d/m_s}$ has been the main reason \cite{Weinberg77}.
To obtain this relation, the texture of the  down-type quark mass matrix must have a zero entry \cite{Weinberg77,Zee77}. However, in the real world with three families, the expansion in terms of  $\sqrt{m_d/m_s}$ is a preposterous suggestion.  Rather, $\halfep\equiv \sqrt{\tt  (mass\, of \, 2nd\, family\, member)/ (mass\, of \, 3rd\, family\, member)}$ should be an expansion parameter \cite{Fritzsch79}.  Since this early study, the texture scenarios of the quark mass matrix  have boomed in the last four decades \cite{QMtexture,Ross18,Davidson84}.  The texture zero entries are assumed to arise from some symmetry. In this paper, we do not follow this line of argument, but take first a general form for the mass matrix such that all the entries are allowed in terms of a small expansion parameter $\halfep$ and next guide to some symmetry based on this phenomenologically determined progenitor mass matrix. If some elements of the CKM matrix are smaller than those of our ansatz, we regard them as possible texture zeros.

\bigskip
\noindent
In general, the CKM matrix, $U^{(u)} U^{(d) \dagger}$, is given by the product of two left-hand (L-hand) unitrary matrices $U^{(u)}$ and $U^{(d)}$ diagonalizing respectively the up-type (\ie $\Qem=+\frac23$ quarks) and  the down-type   mass matrices.  
The matrix  $U^{(u)} U^{(d) \dagger}$ is a unitary matrix, and hence can be written as a unitary matrix $V_{\rm CKM}$. Therefore, defining $V_{\rm CKM}$ with general up-type and down-type weak eigenstates is equivalent to  defining it with a general up-type weak eigenstate and diagonal (or mass)  down-type eigenstates.  
These two cases have the same number of physical parameters. Let us choose this simple case. We can choose the diagonal bases as those of up- or down-type quarks. But, since $m_c/m_t$ is smaller than $m_s/m_b$, we use the bases where the down-type quark mass matrix is already diagonalized \cite{FK19}. 
  In this case,  the CKM matrix $V_{\rm CKM}$ is $U^{(u)} $ which can be expanded in terms of the  small expansion parameter, $\sqrt{\varepsilon}\equiv \sqrt{m_c/m_t}\simeq 0.0584$. Out of 9 parameters in  $U^{(u)} $, two phases can be absorbed to   two up-type quark phases. One overall phase cannot be used for this purpose because of the baryon number  conservation in the SM. The remaining 7 parameters encode three  up-type quark masses, three real angles, and a phase $e^{i\delta}$.
 Instead of texture zeros, we assume the following mass matrix, for its determinant  being of O($\varepsilon^3$)  \cite{FK19},
 \begin{equation}
 M= \left(
\begin{array}{ccc}
 {\rm O(\varepsilon^2)}, & {\rm O(\thalfep)}, &  {\rm O( \oneep)}  \\
{\rm O(\thalfep)}, & {\rm O(\oneep)}, & {\rm O(\halfep)}\\
 {\rm O(\oneep)}, & {\rm O(\halfep)}, & 1 \\
\end{array}
\right).\label{eq:eexpansion}
 \end{equation}
    The electroweak (EW) scale is defined usualy as the $Z^0$ boson pole, but here we define the EW scale as the top mass pole which is the highest scale among SM poles. Then, above the EW scale, we need not consider threshold effects. At the EW scale or at the top mass pole $m_t=172.5$ \cite{CMS19topmass}, $ m_c/m_t|_{m_t}=0.00341\pm 0.00046$ where $m_c=1.25$ GeV of the charm mass  is runned to the EW scale   \cite{Xing07}.  $m_u\simeq 2.5\,$MeV \cite{Manohar18} is run to the electroweak scale to 0.392\,MeV.\footnote{From the QCD scale to the scale  $m_c$, we multiplied a factor 1/3.}
Then,  $m_u\sim m_t\varepsilon^2\simeq 0.001\textrm{\,MeV}=10^{-2}(O(m_t\oneep^2))$ at the top mass pole where  $\varepsilon=0.0584$.

\bigskip
\noindent  
To place zero entries at some places \cite{Ross18}, some kind of symmetry is needed, for example by a kind of the Froggatt--Nielson mechanism \cite{FN79}.  
Theoretically, it is important to know the progenitor mass matrix  determined in this kind of way. So far, there has not appeared any reliable progenitor mass matrix. The essence of this paper is to determine the progenitor form starting from 
Eq. (\ref{eq:eexpansion}), giving correctly all the quark masses by fitting to the experimentally determined CKM matrix elements as accurately as possible.
This progenitor mass matrix can be very useful for future texture studies of the quark mass matrix.

\bigskip
\noindent  
It is better to take bases where both the CKM and the PMNS matrices are defined by the charge raising CCs or by the charge lowering CCs. The currently used convention is that the CKM matrix is defined by the  charge raising CCs but the PMNS matrix by the  charge lowering CCs \cite{PDG18}. It has its own reason that the potential $V$ is important in mixing in the quark sector while the kinetic energy is more important  in the oscillation of   neutrinos. Here, however, we try to describe both the CKM and PMNS matrices in a unified form by the charge raising CCs. Therefore, our PMNS matrix is the hermitian conjugate of that defined in the PDG book \cite{PDG18}. In this paper, however, we pay attention to the CKM matrix only. 
 
\bigskip
\noindent
In Sec. \ref{sec:LRU}, we obtain the up-type quark mass matrix at the EW scale in the bases where the down-type quark mass matrix is already diagonalised. In Sec. \ref{sec:MatGUT}, we run this EW mass matrix to the GUT scale ($\simeq 2.5\times 10^{16\,}\gev$), obtaining  the progenotor quark mass matrix.  In Sec. \ref{sec:Zeros}, we discuss possible zeros in the mass matrix obtained in Secs. \ref{sec:LRU} and \ref{sec:MatGUT}.  Finally, a brief conclusion is given in Sec. \ref{sec:Conclusion}.

\section{Determination of the R-hand Unitary Matrix}\label{sec:LRU}
  
 \bigskip
 \noindent
 We choose the bases where the down-type quarks are mass eigenstates. To unify with leptons, we generalize this to choose  the $T_3=-\frac12$ members in the   L-hand  SM doublets are mass eigenstates, \ie  $\Qem=-1$ leptons and  down-type quarks are mass eigenstates. This choice is simple because the renormalizable couplings are enough for generating the mass terms of these  $T_3=-\frac12$ members in both the quark and lepton sectors. 
Since neutrino masses are not so hierarchical as the up-type quarks, we do not apply the same criteria for the PMNS and neutrino mass matrices.
Even though the CKM and PMNS matrices are of very different forms, they can be successfully unified in GUTs from  string compactification \cite{KimRp19}. An explicit example with the tetrahedral symmetry is already given in Ref. \cite{KimTetra20}.

 \bigskip
 \noindent
 Let us consider a realistic $3\times 3$ matrix,
 \begin{equation}
 M= \left(
\begin{array}{ccc}
 q\varepsilon^2, & f \thalfep, &  c\oneep \\
g \thalfep, & p\oneep, & a\halfep \\
 d \oneep, &  b \halfep, & 1 \\
\end{array}
\right),\label{eq:NumInput}
 \end{equation}
 where $a,b,c,d,f,g,p$ and $q$ are O(1) numbers. Note the possibility that higher order corrections by loops and gravitational interactions can affect the numbers $q,f$ and $g$.  
$M$ is diagonalized by the following  L-hand  matrix $U$,
 \begin{equation}
U= \left(\begin{array}{ccc} c_1,&s_1c_3,& s_1s_3  \\ [0.2em]
-c_2s_1, & c_1c_2c_3+ s_2s_3 e^{-i\delta},&c_1c_2s_3-s_2c_3e^{-i\delta}  
 \\[0.2em]
 -s_1s_2e^{+i\delta},&  -c_2s_3 +c_1s_2c_3 e^{+i\delta}, & c_2c_3 +c_1s_2s_3 e^{+i\delta}
\end{array}\right),\label{eq:Uin}
 \end{equation}
where  $c_i$ and $s_i$ are  cosines and sines of three real angles $\theta_i\,(i=1,2,3)$,
 and the right-hand (R-hand) matrix $W$, by $WMU^\dagger=M^{\rm (diag)}$.   $W$ is determined if the matrix $M$ of the form (\ref{eq:NumInput}) is given. The unitarity condition introduces 9 parameters in  $W$ out of which two phases\footnote{We cannot use the overall phase of the  R-hand matrix  for removing a phase.} can be absorbed to the R-hand up-type quarks. Thus, we can consider the 7 parameter $W$. Since a 7 parameter unitary matrix is not known, we use the following form for $W$ (\ie the same form as  Eq. (\ref{eq:Uin}))
 \begin{equation}
W=  \left(\begin{array}{ccc} {\tt c}_{\bf 1},& {\tt s}_{\bf 1}{\tt c}_{\bf 3},& {\tt s}_{\bf 1}{\tt s}_{\bf 3}  \\ [0.2em]
-{\tt c}_{\bf 2}{\tt s}_{\bf 1}, & {\tt c}_{\bf 1}{\tt c}_{\bf 2}{\tt c}_{\bf 3}+ {\tt s}_{\bf 2}{\tt s}_{\bf 3} e^{-i\alpha},&{\tt c}_{\bf 1}{\tt c}_{\bf 2}{\tt s}_{\bf 3}-{\tt s}_{\bf 2}{\tt c}_{\bf 3}e^{-i\alpha}  
 \\[0.2em]
 -{\tt s}_{\bf 1}{\tt s}_{\bf 2}e^{+i\alpha},&  -{\tt c}_{\bf 2}{\tt s}_{\bf 3} +{\tt c}_{\bf 1}{\tt s}_{\bf 2}{\tt c}_{\bf 3} e^{+i\alpha}, & {\tt c}_{\bf 2}{\tt c}_{\bf 3} +{\tt c}_{\bf 1}{\tt s}_{\bf 2}{\tt s}_{\bf 3} e^{+i\alpha}
\end{array}\right) , \label{eq:URin}
 \end{equation}
where  ${\tt c}_{\bf i}$ and ${\tt s}_{\bf i}$ are  cosines and sines of three real angles $\varphi_{\bf i}\,(\bf i=1,2,3)$.  The angles $\varphi_{\bf i}\,(\bf i=1,2,3)$ will introduce more parameters in them, as  they are expanded in powers of $\halfep$,
 \begin{equation}
\varphi_{\bf i}= \phi_i[0] + \phi_i[1/2] \halfep+ \phi_i[1] \oneep +\cdots
 \end{equation}

\bigskip
\noindent
Let us use the central points of data evaluated by the Kim--Seo (KS) form \cite{KimSeo11} for the CKM  matrix  \cite{KKNS19}, 
 \begin{eqnarray}
V_{\rm CKM}^{\rm KS}&=&
\begin{pmatrix}
 0.975188  , & 0.221345 , & 0.003888 \\[0.5em]
-0.221226 ,  & 0.974365   +0.00065\,e^{-i\delta} , & 0.01712-0.03712\,e^{-i\delta}   \\[0.5em]
  -0.00822\,e^{i\delta} \, , & -0.017551 +0.03620\,
  e^{i\delta} \, ,   &
  0.999156  +0.00064\,e^{i\delta}  
\end{pmatrix}\nonumber  \\[0.3em]
&=&
\begin{pmatrix}
0.975188,&0.221345,&0.555429\,\oneep\\
   -0.221226,&0.974365+1.10986\,e^{-i\delta}\,\thalfep,&(0.204623-0.443669\,e^{-i\delta})\halfep\\
   -1.17429\,e^{i\delta}\,\oneep,&(-0.209775+0.432623\,e^{i\delta})\halfep,&0.999156+1.09278\,e^{i\delta})\thalfep
\end{pmatrix}\label{eq:aCKMdata}
 \end{eqnarray}
which gives $J= (3.114\pm 0.325)\times 10^{-5}|\sin\delta|$ \cite{KimSeo12}.  We scanned $W$ to make $WMU^\dagger$ diagonal to $M^{\rm(diag)}\simeq {\rm diag.}(2.27\times 10^{-6},0.00341,1)={\rm diag.}( 0.195\,\oneep^2,\oneep,1)$.  
Since $M^{\rm(diag)} U$ is ($\oneep=0.00341, \halfep=0.0584$)
 \begin{equation}
\begin{pmatrix}
0.28768 \oneep^2,& 0.0652968  \oneep^2,&0.163516  \oneep^3\\
-0.221156 \oneep,& 0.974365 \oneep+1.10986 e^{-i\delta}\oneep^{5/2},&(0.204623 -0.443669 e^{-i\delta})\oneep^{3/2}\\
-1.17429 e^{i\delta}\oneep ,& (-0.209775+  0.432623 e^{i\delta})\halfep,&0.999156 +1.09278  e^{i\delta}\thalfep
\end{pmatrix},\nonumber
 \end{equation}
which is not close to Eq. (\ref{eq:eexpansion}) or Eq. (\ref{eq:NumInput}), we need an appropriate $W$ to make $M$ match with the form  Eq. (\ref{eq:eexpansion}).   
We find that the real angles of $W$, giving the form (\ref{eq:NumInput}) approximately, are
 \begin{eqnarray}
&&\phi_1=p_1[1]\oneep +\cdots,\nonumber\\
&& \phi_2=p_2[0]+p_2[\frac12]\halfep+p_2[1]\oneep +\cdots,\label{eq:Wfound}\\
&&\phi_3=p_3[\frac12]\halfep +p_3[1]\oneep +\cdots \nonumber 
 \end{eqnarray}
where $p_i[x]$ are  the O(1) coefficient of $\oneep^x$ for the real angle $\varphi_{\bf i}$ in Eq. (\ref{eq:URin}).  There are more than six real numbers ($p_1[1], p_2[0], p_2[\frac12], p_2[1], p_3[\frac12], p_3[1]$) and a phase $e^{i\alpha}$ in our expansion of $W$.
Using $V_{\rm CKM}^{\rm KS}$ as  $U$,  $M=W^\dagger M^{\rm(diag)}U$  becomes for $\alpha=0$,
\begin{equation}
M= 
\left(
\begin{array}{ccc}
 0, & 
 \begin{array}{c}   \Big(0.300555 p_1[1] S[p_2[0]]\\
 -0.619914 e^{i \delta } p_1[1] S [p_2[0]]\Big)\thalfep
 \end{array} , & 
 -0.999156 p_1[1] S [p_2[0]] \oneep\\[2em]
 \begin{array}{c}   \epsilon^{3/2} \Big(0.221226 p_{23}^{-}[\frac{1}{2}] S[p_2[0]]
 \\
  -2.41056 e^{i \delta } p_{23}^{-}(\frac{1}{2}) C[p_2[0]]\Big)\\
  +\epsilon  \Big(-0.221226 C[p_2[0]]\\
 -2.41056 e^{i \delta } S [p_2[0]]\Big)\end{array}, & 
  \begin{array}{c}  \Big(-0.300555 S [p_2[0]]\\
  +0.619914 e^{i \delta} S[p_2[0]]\Big)\halfep\\
  +C[p_2[0]]  \Big( 0.974365+\Big[0.619914 e^{i \delta} \\  
  -0.300555] p_{23}^{-}[\frac{1}{2}\Big] \Big) \oneep \end{array}, & 
   \begin{array}{c}  0.999156 S [p_2[0]]\\
 +0.999156 p_{23}^{-}[\frac{1}{2}] C [p_2[0]]\halfep 
 \end{array} \\[4.5em]
 \begin{array}{c}   \epsilon  \Big(0.221226 S [p_2[0]]\\
 -2.41056 e^{i \delta } C [p_2[0]]\Big)
 \end{array}, & \begin{array}{c}\Big(-0.300555 C[p_2[0]]\\
 +0.619914 e^{i \delta } C [p_2[0]]\Big)\halfep
 \end{array}, & 
 0.999156 C [p_2[0]]   \\
\end{array}
\right)\label{eq:TheForm}
\end{equation}
where  $ p_{23}^{-}=  p_2[\frac12]-p_3[\frac12] ,$ and we kept only the leading terms. $S[p_2[0]]$ cannot be zero  or the determinant is zero. We can obtain an upper bound on  the  $S[p_2[0]]$ such that its contribution to the determinant is less than O($\oneep^3$) from every element of $M$. For a small  $S[p_2[0]]$,   $M_{12}M_{23}M_{31}, M_{13}M_{32}M_{21}, M_{31}M_{22}M_{13}$ and $M_{33}M_{12}M_{21}$ are  O($\oneep^3$),O($ \oneep^{5/2}$ ), O($ \oneep^{5/2}$) and O($ \oneep^{5/2}$ ), respectively.  Therefore, from $M_{13}M_{32}M_{21},M_{31}M_{22}M_{13} $ and $M_{33}M_{12}M_{21} $, we have bounds $S[p_2[0]]<\halfep, 2.4 S[p_2[0]]^2<\halfep,\{0.22 S[p_2[0]], 2.4 S[p_2[0]]^2\}<\halfep$, respectively. These give $S[p_2[0]]<0.0584 ,  0.156,0.265,$ and  0.156, respectively. Thus, we have the common region
\begin{equation}
0<|S[p_2[0]]|<0.0584. \label{eq:S2Bound}
\end{equation}
In a somewhat large region of Eq. (\ref{eq:S2Bound}) for an illustration, we take  $S[p_2[0]]=0.05$  (with $C[p_2[0]]=0.99875$) for which  $M$ is   
\begin{equation}
M\Big([p_2[0]=2.866^{\rm o}]\Big)=
\left(
\begin{array}{ccc}
 0, &
 (0.0150231 -0.030986 e^{i \delta }) p_1[1]\thalfep ,& 
  -0.0499422 p_1[1] \oneep  \\[2em]
\begin{array}{c} (-0.220949-0.12049 e^{i \delta }) \oneep\\
 + \Big(0.0110578 p_{23}(\frac{1}{2})\\
 -2.40754 e^{i \delta } p_{23}\left(\frac{1}{2}\right)\Big) \thalfep \end{array},& 
\begin{array}{c}  (-0.0150231+0.030986 e^{i \delta }) \halfep\\
+ \Big( 0.973147+(0.619139 e^{i \delta} \\
-0.30018) p_{23}[\frac{1}{2}]\Big)\oneep\end{array} ,& 
 \begin{array}{c} 0.0499422 \\
 +0.997907 p_{23}[\frac{1}{2}] \halfep \end{array}
 \\[3em]
  (0.0110578\, -2.40754 e^{i \delta })\oneep ,& \left(-0.30018+0.619139 e^{i \delta }\right) \halfep,& 
 0.997907 
\end{array}
\right)\label{eq:Mew}
\end{equation}
where we used $ \halfep\simeq 0.05840$ at the top mass pole for the (22) element.  
The matrix (\ref{eq:Mew}) is almost the one given in Eq.  (\ref{eq:eexpansion}).  
Indeed,  using Eqs.  (\ref{eq:aCKMdata}) and (\ref{eq:Wfound}), the product  $ W MU^\dagger$ is shown to be diagonal with the error of O($10^{-5}$) for the off-diagonal elements,  
\begin{equation}
WMU^\dagger=
\left(
\begin{array}{ccc}
 2.267 \times 10^{-6} , & -1.499  \times 10^{-10} ,& 0
  \\[1em]
 -2.255  \times 10^{-7}  ,& 
 3.41  \times 10^{-3}-4.334 \times 10^{-6}\cos\delta ,& 1.525  \times 10^{-8}+6.21  \times 10^{-9}  e^{-i \delta }
  \\[1em]
\begin{array}{c} -1.076  \times 10^{-7}\\
-3.356  \times 10^{-6}e^{i \delta} \end{array},&  
\begin{array}{c} 4.471  \times 10^{-6}+1.369  \times 10^{-6} e^{i \delta }\\
-2.268  \times 10^{-7} e^{2 i \delta} 
\end{array} ,& 
 0.9999 +0.8227   \times 10^{-5}  \\
\end{array}
\right)\label{eq:TheFormSp}
\end{equation}

\section{A progenitor form}  \label{sec:MatGUT} 
\noindent
The pole mass is related to the running mass, up to three gluon loops, as
\begin{equation}
M_q=m_q(\mu)\left[ 1+\frac43\frac{\alpha_s(\mu)}{\pi} +K_q^{(2)}\Big(\frac{\alpha_s(\mu)}{\pi} \Big)^2
+K_q^{(3)}\Big(\frac{\alpha_s(\mu)}{\pi} 
\Big)^3 +K_{tt}\delta_{3q}\right] \equiv m_q(\mu)(R_q(\mu)+K_{tt}\delta_{3q})
\end{equation}
where $K_c^{(2)}=11.21, K_t^{(2)}=9.13, K_c^{(3)}=123.8$, and $K_t^{(3)}=80.4 $ are taken from \cite{Xing07}. The running masses at $M_t$  are $m_t(M_t)=162.75$ GeV and  $m_c(M_t)= 554.5$ MeV are found at the top mass pole, using  $\alpha_s=0.108\pm 0.002$ at $\mu=M_t$.

  \bigskip
 \noindent
Above the EW scale or the top mass scale, we can consider the running of $M_{ij}$, considering the gluon loops of Fig.  1 \cite{Chetrykin00}.  
The vector couplings of gluons to quarks allow the same couplings to L-hand and R-hand quarks,  $u_R^{(i)}$ and $u_L^{(j)}$.

\bigskip
\noindent
  The element $M_{ij}$ have the same factor after the running. The reason is the following. The diagonalized mass matrix at $m_t$ is $M^{\rm diag} (m_t)$. The R-hand and L-hand diagonalizing matrix at $m_t$ relate it to the weak eigenstate mass matrix
\begin{equation}
   M(m_t)_{ij} =  \Big(W^\dagger(m_t) M^{\rm diag} (m_t) U(m_t)\Big)_{ij}  .\label{eq:Atmt}
\end{equation}
To obtain a progenitor mass matrix at the GUT scale, each factor in Eq. (\ref{eq:Atmt}) is runned to the GUT scale. For the diagonal masses, the knowledge on the anomalous dimension suffices. For the R- and L-hand unitary matrices, strong interaction corrections do not distinguish them because the perturbative QCD conserves parity. The $\times$ in Fig. 1 corresponds to $ M (\mu)_{ij}  $ and every gluon  vertex  on the fermion line has the same 1-loop factor $(1+a_1\alpha_s)$ because it is flavor blind. Therefore, we can take out $(1+a_1\alpha_s)$ at 1-loop level
\begin{equation}
 (1+a_1\alpha_s(\mu))^2  \Big(W^\dagger(m_t) M^{\rm diag} (\mu) U(m_t)\Big)_{ij} = W^\dagger(\mu)  M (\mu)_{ij}U(\mu)  = R(\mu) M_{ij}(m_t).
\end{equation}
Taking a ratio ${M(\mu)_{ij}}/{M(\mu)_{33}}$, we obtain
\begin{eqnarray}
 \frac{M(\mu)_{ij}}{M(\mu)_{33}}= \frac{ \Big(W^\dagger(m_t) M^{\rm diag} (\mu) U(m_t)\Big)_{ij} }{ \Big(W^\dagger(m_t) M^{\rm diag} (\mu) U(m_t)\Big)_{33} }.
\end{eqnarray}
where the flavor independence of gluon couplings is used. Therefore, the progenitor mass matrix is proportional to Eq. (\ref{eq:Mew}), except the (33) element.   Let
$M_{ij}(\mu)= R(\mu) M_{ij}(m_t)$ where $R(m_t)=1$, but  the (33) element may be modified significantly due to a large top quark Yukawa coupling constant.  
The Higgs loop for is considered as $K_{tt}=-\frac{3}{16\pi^2}(m_t^2/v_u^2)$ where the Higgs doublet  $H_u$ couples to the top quark as
\begin{equation}
  \frac{m_t}{\sqrt2 v_u} (H_u^\dagger) \bar{t}_R q_{3L}+h.c.,\\[0.5em]
   H_u=\begin{pmatrix}
  H_u^+\\[0.4em]
   \sqrt2 v_u +\frac{h_u^0}{\sqrt{2}}
  \end{pmatrix}.
\end{equation} 
If $H_u$ is the only electroweak Higgs doublet, the (33) element runs to the  scale $\Mg\simeq 2.5\times 10^{16}$ GeV to  $R_t(\Mg)=1.01123$ for $\alpha_s(\Mg)\simeq \frac{1}{40}$, and $K_{tt}(\Mg)=- 0.0094$. So, the (33) element changes factor 1 to  factor 0.9703. Even if we added the top quark Yukawa coupling, the progenitor form obtained from Eq. (\ref{eq:Mew}) is almost intact. Now, we study the form (\ref{eq:TheForm}) from the point of view of introducing texture zeros.
\begin{figure}[!t]\label{fig:Running}
\hskip 0.01cm \includegraphics[width=0.95\textwidth]{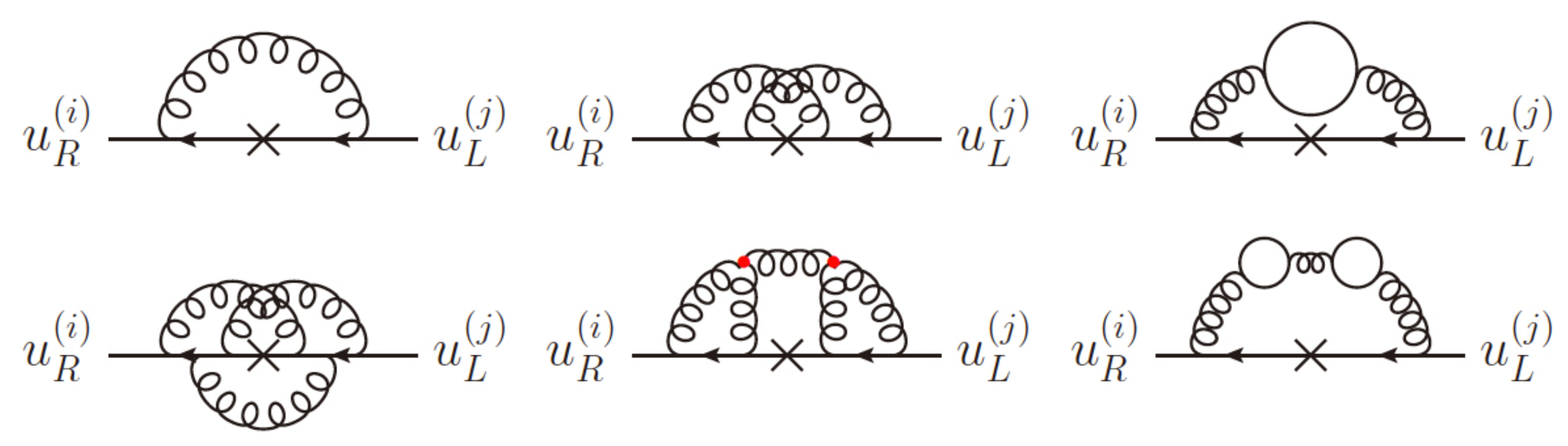}
\caption{Some diagrams contributing the the up-type quark propagators \cite{Chetrykin00}. Here, $\times$ are the elements $M_{ij}$.}
\end{figure}

\section{Texture}\label{sec:Zeros} 
 
\noindent  
Since $S[p_2[0]]$ is very small, we can consider $M_{12}$ being zero in the first approximation. But we keep  $M_{13}$ as a nonzero value to have a nonvanishing determinant.
Comparing the experimentally determined $M$ with the ansatz form, Eq. (\ref{eq:eexpansion}),  we note that a texture zero element is possible  for the (11)  and (12) elements,
\begin{equation}
M= 
\begin{pmatrix}
 0, & 0, &   {\rm O(\oneep)} \\[0.3em]
{\rm O(\thalfep)}, & {\rm O(\oneep)}, & {\rm O(\halfep)}
\\[0.3em]
 {\rm O(\oneep)}, & {\rm O(\halfep)}, & 1 
\end{pmatrix}\label{eq:Texture0}
\end{equation}
The zeros  in the mass matrix determined from Eq. (\ref{eq:Texture0}) signal  the long-awaited texture form for the quark mass matrix. It is surprisingly simple that the mass matrix determines all the quark flavor parameters. 
 \begin{table}[!h] 
\caption{U(1) and $P$ quantum numbers of the quark and Higgs fields.}
\begin{tabular}{c|ccc|cccc|cccc|ccc} 
\hline
 &&&&&&&&&&&\\[-1.1em]  
 Fields~~& ~$\bar{u}_R^{(1)}$
 & $\bar{u}_R^{(2)}$& $\bar{u}_R^{(3)}$& ~~$q^{(1)}_L$ & $q^{(2)}_L$ &$q^{(3)}_L$ & $H_u$~&~ $\bar{d}_R$ & $\bar{s}_R$ & $\bar{b}_R$& $H_d$&~ $\sigma'$ &~$S'$ \\[0.2em] 
\hline
   $U(1)$~ & $-3$&$-2$  & $-1$& $-1$& $0$& $+1$& $0$& $+1$& $0$& $-1$& $0$& ~$+1$& ~$0$ \\[0.2em] 
\hline
   $P$~ & $- $&$+$  & $+$& $+$& $+$& $+ $& $+$& $-$& $+$& $+$& $+$& ~$+ $& ~$-$ \\
    \hline
  \end{tabular}\label{tab:U1s} 
\end{table}
In Table \ref{tab:U1s}, we present U(1)  times parity $P$ quantum numbers to have the form Eq. (\ref{eq:Texture0}).
The L- and R-states are defined as
\begin{equation}
\bar{u}_R^{(i)}  M_{ij}  q_L^{(j)}
=\bar{u}_R W^\dagger WM U^\dagger U q_L 
=\bar{u}_R^{\rm mass} M^{\rm diag}   q_L^{\rm mass}
\end{equation}
where $\bar{u}_R$ and $u_L$ are in the weak bases, and $u_R^{\rm mass}$ and $u_L^{\rm mass}$  are in the mass eigenstates, e.g. $u_L^{\rm mass}=Uu_L$, etc. $\bar{d}_R, \bar{s}_R$ and $\bar{d}_R$ are the mass eigenstates, $d_R^{ (j)\,\rm mass}$, and 
\begin{equation}
q_L^{(j)}=\begin{pmatrix}{u}_L^{(j)}\\[0.5em]
 d_L^{ (j)\,\rm mass}
\end{pmatrix}.
\end{equation}
 Table  \ref{tab:U1s} allows the following mass matrix
\begin{equation}
M=m_t\begin{pmatrix}
 q\,\sigma^4S,&  f\,\sigma^3S,& c\, \sigma^2S\\
 g\,\sigma^3,&  p\,\sigma^2,&a\,\sigma
 \\
d\,\sigma^2,& b\,\sigma,&1
\end{pmatrix}
\end{equation}
where $a,b,c,d,f,g,p,$ and $q$ are O(1) numbers, and $\sigma$ and $S$ are dimensionless fields defined at the scale $\Mg$, $\sigma=\frac{\sigma'}{\Mg}$ and  $S=\frac{S'}{\Mg}$.

\section{Conclusion}\label{sec:Conclusion} 
\noindent
We determined the quark mass matrix in terms of a small expansion parameter $\halfep$, which gives correctly all the quark masses and the CKM matrix elements at the  EW scale, and obtain a progenitor form at the GUT scale by running this EW scale mass matrix.  Finally, a possible texture form for the progenitor quark mass matrix is suggested.
  
\acknowledgments{We thank Sin Kyu Kang for helpful discussions.  This work is supported in part  by the National Research Foundation (NRF) grant  NRF-2018R1A2A3074631.}

 
 \end{document}